\documentclass[prx,aps,twocolumn,superscriptaddress]{revtex4-2}

\usepackage[pdftex]{graphicx}
\usepackage{amsbsy,amssymb,amsmath,bm,mathtools}
\usepackage{amsfonts}

\usepackage{xspace}
\usepackage{bm}
\usepackage{color}
 
\usepackage{url}
\usepackage[section]{placeins}
\usepackage[colorlinks=true,linkcolor=blue,citecolor=blue]{hyperref}

\usepackage[mathscr]{euscript}

\usepackage{graphicx}
\usepackage{dcolumn}

\begin{document}

\preprint{APS/123-QED}

\title{Graph Neural Network Force Fields for Spin Dynamics in Metallic Magnets}

\author{Ali Rayat}
\affiliation{Department of Physics, University of Virginia, Charlottesville, VA 22904, USA}

\author{Yunhao Fan}
\affiliation{Department of Physics, University of Virginia, Charlottesville, VA 22904, USA}

\author{Gia-Wei Chern}
\affiliation{Department of Physics, University of Virginia, Charlottesville, VA 22904, USA}

\begin{abstract}
Metallic magnets exhibit complex spin dynamics governed by electronically generated interactions. Predictive simulations of such dynamics typically require repeated solutions of an underlying electronic problem throughout the time evolution, creating a major computational bottleneck. Here we introduce a graph neural network (GNN) magnetic force-field framework that learns the effective magnetic energy functional governing itinerant spin dynamics directly from electronic calculations. Conceptually analogous to machine-learned interatomic potentials, the proposed framework enables efficient evaluation of spin torques while capturing the nonlinear and spatially extended interactions generated by itinerant electrons. We benchmark the method on representative metallic magnetic systems exhibiting collinear, noncollinear, and noncoplanar magnetic order. The learned force fields accurately reproduce electronically generated spin torques and yield nonequilibrium spin dynamics in excellent agreement with direct electronic simulations. Our results establish graph neural networks as a powerful framework for machine-learned magnetic force fields, providing a pathway toward predictive large-scale simulations of nonequilibrium magnetism across multiple length and time scales.
\end{abstract}

\maketitle

\section{Introduction}

\label{sec:intro}

Metallic magnets exhibit a rich interplay between itinerant electrons and localized magnetic moments, giving rise to diverse phases and functionalities ranging from conventional ferro- and antiferromagnets to complex spin textures and spintronic phenomena~\cite{nagaosa13,Zutic2004,Chappert2007,jungwirth16,Tokura2017}. Electronically mediated magnetic interactions, together with crystal structure and lattice geometry, stabilize spin spirals, stripe phases, chiral magnetic orders, skyrmions, and other emergent magnetic textures~\cite{bogdanov89,rossler06,muhlbaure09,yu11}. Such phenomena are central to many functional magnetic materials, including transition-metal ferromagnets, manganites, B20 chiral magnets, kagome metals, and frustrated itinerant magnets. Understanding their nonequilibrium dynamics is therefore a central challenge in magnetic materials research, driven by both fundamental interest and emerging spintronic applications.

Achieving predictive simulations of such dynamics, however, remains computationally demanding despite significant advances in atomistic spin-dynamics methods~\cite{Evans2014,Eriksson2017}. In metallic magnets, the magnetic moments evolve on timescales much slower than the underlying electrons, allowing the electronic subsystem to remain approximately equilibrated with the instantaneous spin configuration. The resulting electronic response determines the effective magnetic fields governing the spin dynamics. Because these fields depend sensitively on the evolving spin configuration, they must be recomputed throughout the time evolution, making repeated solutions of the underlying electronic problem the dominant computational bottleneck. This computational structure closely resembles that of \textit{ab initio} molecular dynamics (MD), where interatomic forces are repeatedly evaluated from electronic-structure calculations during the evolution of atomic coordinates~\cite{Car1985,Marx2009}.

This computational bottleneck naturally motivates the development of the magnetic analogue of machine-learning (ML) force fields~\cite{behler07,bartok10,li15,shapeev16,botu17,smith17,zhang18,Lubbers2018,deringer19,chmiela17,chmiela18,sauceda20}. Over the past decade, ML force fields have transformed atomistic simulations in quantum chemistry and materials science by replacing expensive electronic-structure calculations with neural-network surrogates capable of predicting interatomic forces with near-\emph{ab initio} accuracy at a fraction of the computational cost. This breakthrough has enabled MD simulations on previously inaccessible length and time scales. More recently, the force-field paradigm has begun to extend beyond atomic coordinates to condensed-matter systems with coupled electronic and collective degrees of freedom~\cite{Ma19,zhang22,zhang22b,cheng23a,Liu22,Tian23,Ghosh24,Fan2026}, highlighting the broader applicability of machine-learned force-field methodologies beyond atomistic simulations.

Recent years have also witnessed growing efforts to develop ML force fields for magnetic systems. Conventional approaches extend atomistic force fields to magnetic materials by incorporating local spin degrees of freedom into interatomic potentials~\cite{Eckhoff2021,Kotykhov2024,Yu2024,Yu2024b,Huang2025,Xu2025}, while more recent studies have learned magnetic energy functionals, effective fields, or spin torques directly from electronic calculations to construct surrogate models for spin dynamics~\cite{zhang21,zhang23,cheng23b,Fan24,tyberg25,Chern2026,Chern2026b}. This approach is particularly attractive for itinerant magnets, where the effective interactions responsible for spin dynamics are electronically generated, long-ranged, highly nonlinear, and strongly dependent on the instantaneous spin configuration. By replacing repeated electronic calculations with efficient learned surrogates, machine-learned magnetic force fields provide a promising route toward large-scale simulations of nonequilibrium spin dynamics in metallic materials.

A central challenge in constructing such force fields is the faithful incorporation of the underlying physical symmetries. In conventional molecular dynamics, force fields must respect the Euclidean symmetry group $E(3)$ together with permutation symmetry among chemically identical atoms. Magnetic systems, however, possess a fundamentally different symmetry structure. In the absence of significant spin-orbit coupling, the dynamics is invariant under global spin rotations, while the crystal lattice imposes additional discrete point-group symmetries. Embedding these symmetry constraints into the model is essential for ensuring physically meaningful predictions and robust transferability. Recent studies have shown that symmetry-aware descriptors combined with neural-network force fields can successfully capture these constraints, enabling accurate and scalable surrogate models for nonequilibrium spin dynamics~\cite{behler11,ghiringhelli15,bartok13,himanen20,huo22,drautz19}. However, descriptor-based approaches generally rely on manually constructed features whose expressive power may be difficult to systematically improve as the complexity of the magnetic environment increases, particularly for itinerant magnets with long-ranged, nonlinear, and configuration-dependent interactions.

Graph neural networks (GNNs) provide a natural alternative framework for machine-learned force fields~\cite{scarselli2009,gilmer2017,hamilton2017,xu2019,maron2019,xie2018,schutt2018,choudhary2021,dai2021,reiser2022,batatia2022}. By representing physical systems as graphs and learning interactions through iterative message passing, GNNs construct many-body representations directly from data, allowing increasingly nonlocal and nonlinear interactions to be learned without relying on fixed descriptors. More recently, equivariant neural-network architectures have further expanded the ability of ML force fields to incorporate physical symmetries directly at the architectural level~\cite{cohen2016,cohen2018,weiler2018,kondor2025,batzner2022,kaba2022,musaelian2023,gong2023,batatia2025,yang2025}. Together, graph-based and symmetry-preserving architectures have enabled force-field models that achieve near-\emph{ab initio} accuracy for large-scale molecular and materials simulations while faithfully respecting the underlying physical symmetries.

Motivated by these developments, we develop a GNN force-field framework for itinerant spin dynamics. Rather than adapting architectures originally developed for atomistic simulations, our approach is tailored to the symmetry structure and emergent electronic physics of metallic magnets. The graph is constructed from spin-rotation-invariant descriptors defined on bonds and elementary triangular plaquettes, allowing the network to encode both conventional spin correlations and higher-order geometric information associated with noncoplanar magnetic textures. In particular, the node features incorporate local measures of spin chirality and Berry-phase effects, which play a central role in the coupling between itinerant electrons and complex magnetic backgrounds. Combined with symmetry-preserving message-passing operations, this formulation provides a physically motivated representation of the magnetic environment and enables the network to learn the resulting many-body interactions directly from electronic calculations.

Although the present work employs the itinerant $s$--$d$ model as a controlled benchmark, the proposed framework is formulated independently of the particular electronic solver used to generate the effective magnetic energy functional. As improved microscopic theories of nonequilibrium spin dynamics become available, the same machine-learning framework can be naturally extended without changing its underlying philosophy.

We benchmark the proposed framework on representative itinerant magnetic systems exhibiting collinear, noncollinear, and noncoplanar magnetic orders, including the N\'eel state on the square lattice, the $120^\circ$ state on the triangular lattice, and the noncoplanar tetrahedral phase. The learned force field accurately reproduces electronically generated spin torques and yields Landau--Lifshitz simulations in excellent agreement with direct electronic calculations. Compared with earlier descriptor-based approaches, the proposed GNN framework combines physically motivated magnetic descriptors with graph-based message passing to provide a more flexible and systematically improvable representation of complex magnetic environments while maintaining high predictive accuracy. More broadly, our results establish graph neural networks as a powerful framework for machine-learned magnetic force fields and provide a pathway toward predictive, large-scale simulations of nonequilibrium magnetism across multiple length and time scales.

\section{Graph Neural Network Force-Field Framework}

\label{sec:framework}

We begin by introducing the general GNN force-field framework before specializing to particular microscopic Hamiltonians and magnetic orders. The methodology consists of three main components: a symmetry-preserving initialization of magnetic descriptors, graph-based message passing that progressively builds many-body representations of the magnetic environment, and an energy-based readout from which effective fields and spin torques are obtained through automatic differentiation.
Our focus is itinerant magnetic systems in which the spin dynamics is governed primarily by electronically mediated exchange interactions, while spin-orbit-induced anisotropies remain weak or subdominant. In this regime, the dynamics approximately preserves a global SO(3) spin-rotation symmetry together with the discrete symmetries of the underlying crystal lattice. This symmetry structure arises in a broad class of itinerant-electron models, including double-exchange, Kondo-lattice, and itinerant Heisenberg-type systems, and is directly relevant to many metallic magnets, such as weak-anisotropy \(3d\) transition-metal compounds, manganites, rare-earth intermetallics, and frustrated metallic magnets on triangular and kagome lattices. In these materials, complex magnetic textures often emerge from competing electronically mediated exchange interactions rather than strong spin-orbit coupling.

A particularly important feature of itinerant magnets is that the effective interactions between localized moments are not fixed microscopically, but instead emerge dynamically through their coupling to itinerant electrons. As a result, the magnetic interactions are generally long-ranged, highly nonlinear, and strongly dependent on the instantaneous spin configuration itself. These electronically generated interactions can stabilize a diverse range of complex magnetic orders, including spiral states, multi-\(Q\) textures, and noncoplanar chiral phases. Capturing such emergent many-body interactions within a scalable dynamical framework therefore represents a central challenge for large-scale simulations of itinerant spin dynamics. The GNN framework developed below addresses this challenge by combining physically motivated local magnetic descriptors with graph-based message passing, allowing increasingly nonlocal magnetic interactions to be learned directly from electronic data.

\subsection{Message passing in magnetic environments}

\begin{figure*}[t]
\centering
\includegraphics[width=1.99\columnwidth]{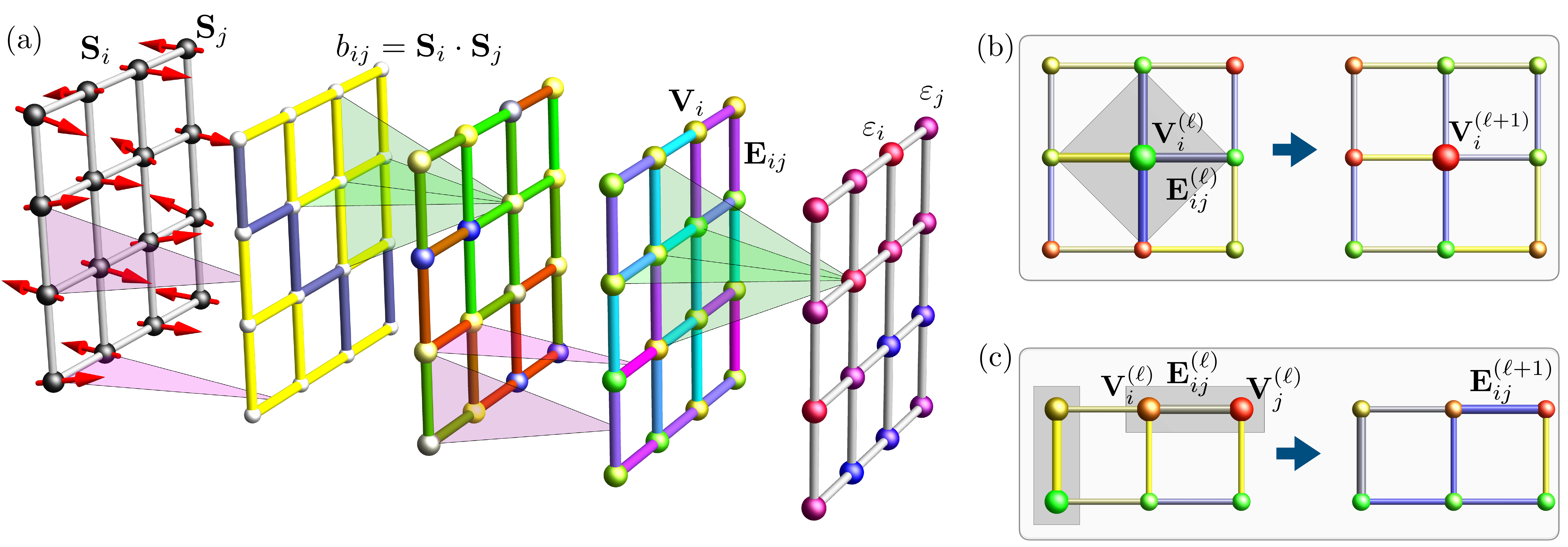}
\caption{
Schematic illustration of the graph-neural-network architecture. 
(a) The magnetic lattice is represented as a graph whose node and edge embeddings are initialized from symmetry-preserving spin features and subsequently refined through successive graph-neural-network layers. 
(b) Node-update operation, in which information from neighboring sites and connecting bonds is aggregated to generate an updated node embedding. 
(c) Edge-update operation, in which the bond embedding is updated using the existing edge feature together with information from the two endpoint sites.
}
    \label{fig:GNN-scheme}
\end{figure*}

The overall GNN architecture is illustrated schematically in Fig.~\ref{fig:GNN-scheme}. The central idea is to construct increasingly expressive representations of the magnetic environment through successive message-passing layers. Starting from local symmetry-preserving descriptors, information is progressively propagated across the graph so that the learned node and edge embeddings encode magnetic correlations extending over multiple coordination shells.

The magnetic lattice is represented as a graph in which lattice sites correspond to nodes and exchange bonds correspond to edges. The central objective is to construct a neural-network force field that preserves the approximate global spin-rotation symmetry of itinerant magnetic systems while simultaneously incorporating the discrete lattice symmetries of the underlying crystal structure.
At message-passing layer $\ell$, each lattice site $i$ is associated with a node embedding $\mathbf{V}_i^{(\ell)}\in\mathbb{R}^{N_V}$, whose components are denoted by $V_{i,\alpha}^{(\ell)}$ with $\alpha=1,\ldots,N_V$. Similarly, each nearest-neighbor bond $(ij)$ is associated with an edge embedding $\mathbf{E}_{ij}^{(\ell)}\in\mathbb{R}^{N_E}$, whose components are denoted by $E_{ij,\beta}^{(\ell)}$ with $\beta=1,\ldots,N_E$. Here $N_V$ and $N_E$ denote the numbers of node and edge feature channels. These embeddings provide latent representations of the local magnetic environment and are iteratively refined through graph message passing.

Following feature initialization, node and edge embeddings are updated through message passing designed to preserve the point-group symmetry of the underlying lattice. At each layer, information is exchanged between neighboring sites and bonds, enabling progressively larger magnetic environments to be incorporated into the learned representation. The resulting aggregated message vector for note-$i$ is
\begin{align}
	\mathbf{M}_i^{(\ell+1)} &= \sum_r \mathbf{W}_{\rm V\to V}(r) \sum_{j\in\mathcal N_i(r)} \alpha_j \mathbf{V}_j^{(\ell)} \nonumber\\
	&\quad+ \sum_r \mathbf{W}_{\rm E\to V}(r) \sum_{j\in\mathcal N_i(r)} \beta_{ij} \mathbf{E}_{ij}^{(\ell)},
	\label{eq:node_message}
\end{align}
where $\mathcal N_i(r)$ denotes the set of neighboring sites belonging to coordination shell $r$. The shell index includes the on-site shell $r=0$, for which $\mathcal N_i(0)=\{i\}$. Consequently, the $r=0$ contribution of the first summation corresponds to the self-update of the node embedding, while the terms with $r>0$ incorporate information from neighboring sites. The scalar coefficients $\alpha_j$ and $\beta_{ij}$ are learned gating factors generated from the local node and edge embeddings, respectively. For example, these coefficients may be obtained from small multilayer perceptrons according to
\begin{equation}
	\alpha_j = \phi_{\rm V}\bigl(\mathbf V_j^{(\ell)}\bigr), \qquad
	\beta_{ij} = \phi_{\rm E}\bigl(\mathbf E_{ij}^{(\ell)}\bigr).
	\label{eq:gating}
\end{equation}
These gates enable the network to assign environment-dependent weights to different neighboring sites and bonds, thereby increasing the expressive power of the message-passing procedure.

The matrices $\mathbf{W}_{\rm V\to V}(r)$ and $\mathbf{W}_{\rm E\to V}(r)$ are trainable weight matrices that determine how node and edge feature channels contribute to the updated node representation. Distinct weight matrices are assigned to different coordination shells, allowing the network to learn shell-dependent interactions while preserving the symmetry equivalence of neighbors within the same shell. Because the aggregation is performed over all sites belonging to a given shell, the resulting update is invariant under point-group operations that permute symmetry-equivalent sites and bonds. For notational simplicity, the layer dependence of the trainable parameters and gating networks has been suppressed.

The aggregated message vector is subsequently combined with a trainable bias vector and transformed by a nonlinear activation function to produce the updated node embedding,
\begin{equation}
	\mathbf{V}_i^{(\ell+1)} = \sigma\!\left( \mathbf{M}_i^{(\ell+1)} + \mathbf b_{\rm V} \right),
	\label{eq:node_activation}
\end{equation}
where $\mathbf b_{\rm V}$ is a trainable bias parameter and $\sigma(\cdot)$ denotes a nonlinear activation function. The activation function is applied component-wise to each feature channel. The introduction of nonlinearity enables the network to represent the highly nontrivial many-body correlations underlying the magnetic energy functional, which cannot be captured by a purely linear sequence of message-passing operations.

The edge embeddings are updated simultaneously with the node embeddings. In contrast to the node update, which aggregates information from multiple coordination shells, the edge update is entirely local and involves only the bond $(ij)$ and the two lattice sites connected by that bond. The corresponding aggregated edge message is
\begin{equation}
	\mathbf N_{ij}^{(\ell+1)} = \mathbf W_{\rm E\to E} \, \mathbf E_{ij}^{(\ell)} 
	+ \mathbf W_{\rm V\to E} \left( \mathbf V_i^{(\ell)} + \mathbf V_j^{(\ell)} \right),
	\label{eq:edge_message}
\end{equation}
where $\mathbf W_{\rm E\to E}$ and $\mathbf W_{\rm V\to E}$ are trainable weight matrices that mix edge and node feature channels, respectively. The first term propagates information already stored in the bond embedding, while the second term incorporates information from the node embeddings located at the two endpoints of the bond. The symmetric combination $\mathbf V_i+\mathbf V_j$ ensures that the edge update remains invariant under interchange of the two sites connected by the bond.

The updated edge embedding is then obtained through
\begin{equation}
	\mathbf E_{ij}^{(\ell+1)} = \sigma\!\left( \mathbf N_{ij}^{(\ell+1)} + \mathbf b_{\rm E} \right),
	\label{eq:edge_activation}
\end{equation}
where $\mathbf b_{\rm E}$ is a trainable bias vector and $\sigma(\cdot)$ denotes a nonlinear activation function.

Repeated application of the message-passing procedure enables information to propagate across progressively larger regions of the lattice. Consequently, the node and edge embeddings become progressively enriched with information about the surrounding magnetic environment. After $L$ message-passing layers, the node and edge embeddings encode many-body spin correlations extending over multiple coordination shells. In this way, the GNN progressively constructs increasingly nonlocal representations of the magnetic environment while preserving the underlying symmetry constraints.

\begin{figure*}[t]
\centering
\includegraphics[width=1.99\columnwidth]{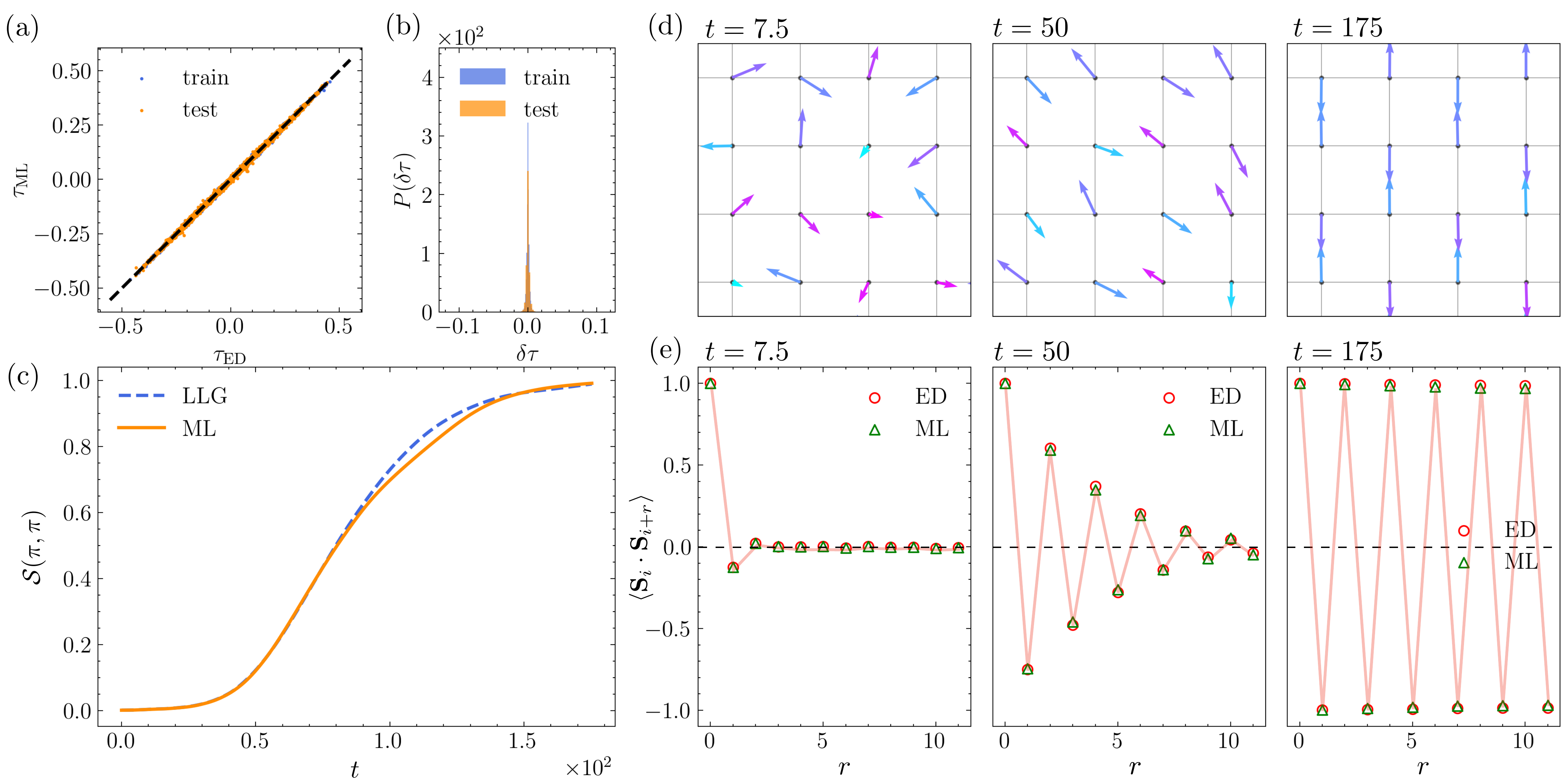}
\caption{
Benchmark of the GNN force field for the square-lattice N\'eel antiferromagnet. (a)~Comparison of torques $\bm\tau_i$ predicted by the GNN and exact electronic calculations. (b)~Corresponding distribution of torque errors. (c)~Time evolution of the spin structure factor evaluated at the N\'eel ordering wave vector $(\pi,\pi)$ following a thermal quench. (d)~Representative snapshots of the spin orientations at different stages of the relaxation dynamics. (e) Spin-spin correlation functions at selected times, comparing direct ED-LLG simulations with GNN-LLG results. The excellent agreement demonstrates that the learned force field accurately reproduces both the local torques and the emergent nonequilibrium dynamics.
}
    \label{fig:neel-benchmark}
\end{figure*}

\begin{figure*}[t]
\centering
\includegraphics[width=1.99\columnwidth]{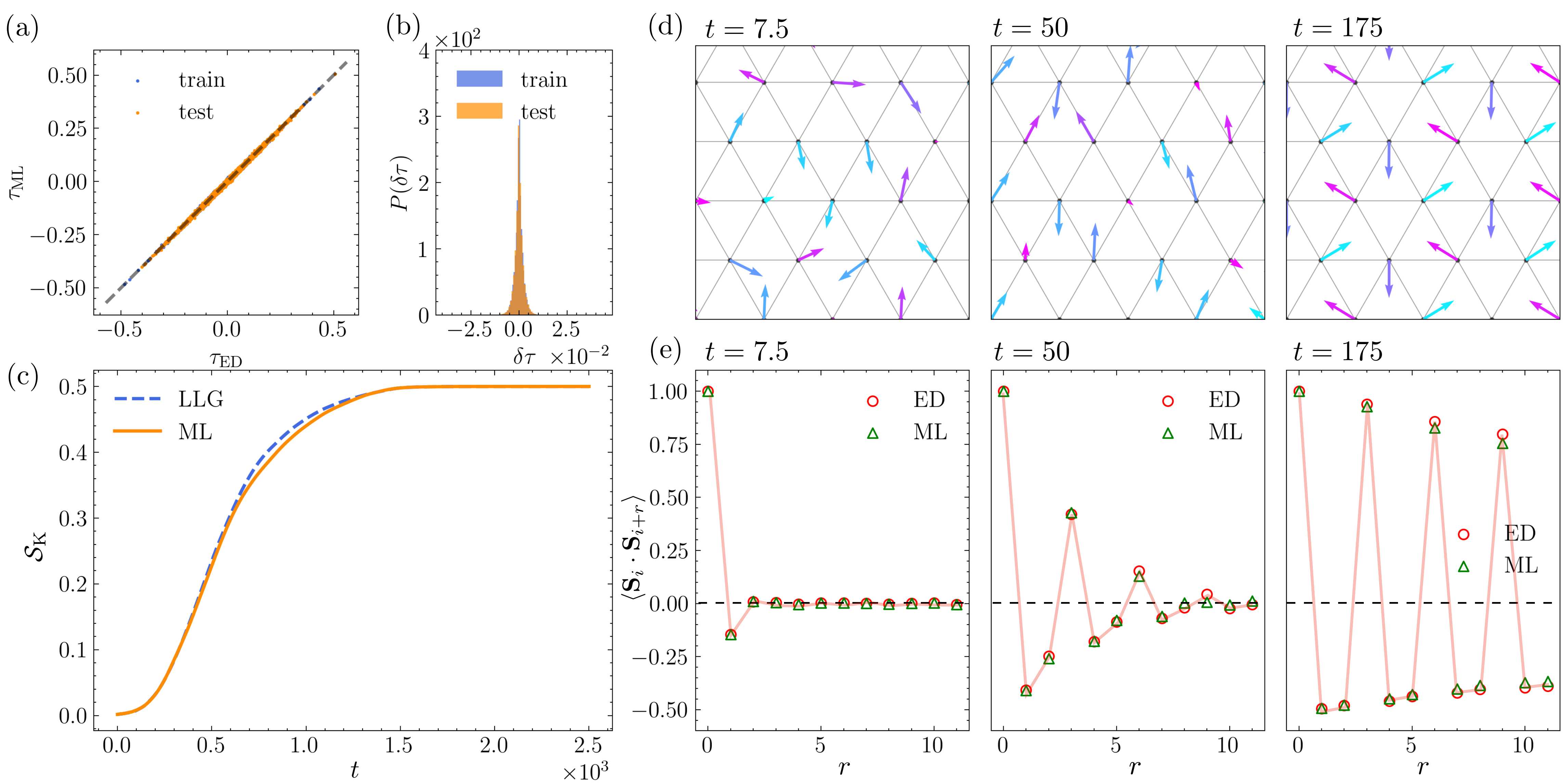}
\caption{
Benchmark of the graph-neural-network force field for the triangular-lattice $120^\circ$ antiferromagnetic state. (a)~Comparison of torques predicted by the GNN and exact electronic calculations. (b)~Corresponding distribution of torque errors. (c)~Time evolution of the spin structure factor $S_K$ at the $K$ point of the hexagonal Brillouin zone, which serves as the instantaneous order parameter for the $120^\circ$ state, following a thermal quench. (d)~Representative snapshots of the spin orientations during the relaxation process. (e)~Equal-time spin-spin correlation functions at selected times, comparing direct ED--LLG simulations with GNN--LLG results. The excellent agreement demonstrates that the learned force field accurately reproduces both the local torques and the nonequilibrium spin dynamics.
}
    \label{fig:120-benchmark}
\end{figure*}

\subsection{Symmetry-preserving feature initialization}

The performance of a graph neural network depends critically on how the local magnetic environment is encoded at the input layer. In the present work, the initial node and edge features are constructed from quantities that respect the fundamental symmetries of the magnetic Hamiltonian. In particular, all input features are invariant under global spin rotations and time reversal, ensuring that these symmetries are preserved from the outset of the message-passing procedure.

The simplest quantity satisfying these requirements is the pairwise spin correlation. We therefore initialize each edge feature according to
\begin{equation}
	\mathbf{E}_{ij}^{(0)} = b_{ij} = \mathbf{S}_i \cdot \mathbf{S}_j,
\end{equation}
which measures the relative alignment of neighboring spins and serves as the fundamental bond feature throughout the graph.

To characterize higher-order spin correlations, we additionally introduce two three-spin quantities associated with elementary triangular plaquettes,
\begin{eqnarray}
	& & \chi_{ijk} = \mathbf{S}_i \cdot \left( \mathbf{S}_j \times \mathbf{S}_k \right), \\
	& & \Lambda_{ijk} = \mathbf{S}_i \cdot \mathbf{S}_j + \mathbf{S}_j \cdot \mathbf{S}_k + \mathbf{S}_k \cdot \mathbf{S}_i. \nonumber 
\end{eqnarray}
The scalar chirality $\chi_{ijk}$ measures the noncoplanarity of the local spin texture, while $\Lambda_{ijk}$ characterizes the overall pairwise spin alignment within the plaquette. Both quantities are invariant under global spin rotations. Under time reversal, however, $\chi_{ijk}$ changes sign whereas $\Lambda_{ijk}$ remains unchanged. To preserve time-reversal symmetry at the level of the input representation, we therefore use the invariant combination $\chi_{ijk}^{\,2}$ together with $\Lambda_{ijk}$ as the elementary plaquette features.

The initial node embedding is constructed by collecting the plaquette features associated with all triangles sharing the vertex $i$,
\begin{eqnarray}
	\mathbf{V}_i^{(0)} = \Big( \chi_{ijk}^{\,2}, \chi_{ij'k'}^{\,2}, \ldots, \Lambda_{ijk}, \Lambda_{ij'k'}, \ldots \Big), 
\end{eqnarray}
where each entry corresponds to a distinct triangular plaquette containing site $i$. This representation preserves detailed information about the local magnetic geometry while satisfying the exact symmetry constraints of the problem. In contrast to approaches that compress local information into a small number of aggregated descriptors, the present construction retains the individual plaquette contributions and allows the network to learn their optimal combinations during message passing.

The inclusion of these plaquette-based features substantially improves the predictive accuracy of the model compared with using pairwise spin correlations alone, demonstrating the importance of higher-order geometric information in describing itinerant magnetic systems. Beyond their favorable empirical performance, the quantities $\chi_{ijk}$ and $\Lambda_{ijk}$ also possess a clear physical interpretation. In the strong-coupling limit, the Berry phase $\Phi_{ijk}$ accumulated by an electron traversing a triangular path in a noncoplanar spin texture is determined by both quantities through the relation $ \tan\Phi_{ijk} = -\chi_{ijk}/(1+\Lambda_{ijk})$.  The quantities $\chi_{ijk}^2$ and $\Lambda_{ijk}$ therefore provide physically motivated descriptors of the local magnetic geometry and emergent gauge structure experienced by itinerant electrons, furnishing symmetry-preserving features from which the GNN learns the electronically generated many-body interactions.

\subsection{Energy readout and Landau--Lifshitz dynamics}

The final node and edge embeddings produced by the message-passing procedure provide flexible representations from which a variety of physically relevant quantities may be constructed. The final node and edge embeddings provide flexible latent representations from which a variety of physical observables may be predicted. In the present work, however, we adopt a site-centered energy decomposition analogous to the Behler--Parrinello framework and use only the final node embeddings $\mathbf{V}_i^{(L)}$ in the readout stage. These node representations are subsequently mapped onto local energy contributions through a neural-network readout function. Specifically, each final node embedding is processed by a multilayer perceptron (MLP) $\psi(\cdot)$ according to
\begin{equation}
	\epsilon_i=\psi\bigl(\mathbf{V}_i^{(L)}\bigr),
\end{equation}
where the same network $\psi$ is applied to all lattice sites. The total magnetic energy is then obtained as
\begin{equation}
	E=\sum_i \epsilon_i.
\end{equation}
Although $\epsilon_i$ is commonly referred to as a local energy, it should be regarded as a latent variable rather than a uniquely defined physical observable. Because the node embeddings have accumulated information from neighboring sites through multiple message-passing layers, each contribution generally depends on a finite magnetic environment surrounding site $i$. Nevertheless, the decomposition above provides an efficient and symmetry-preserving representation of the magnetic energy functional.

The effective magnetic field acting on spin $\mathbf{S}_i$ is obtained directly from the learned energy functional through automatic differentiation,
\begin{equation}
	\mathbf{H}_i=-\frac{\partial E}{\partial \mathbf{S}_i}.
\end{equation}
Deriving the fields from a scalar energy functional ensures the internal consistency of the predicted torques and guarantees that the resulting dynamics evolves on a well-defined magnetic energy landscape.

The spin dynamics is subsequently governed by the stochastic Landau--Lifshitz--Gilbert (LLG) equation,
\begin{equation}
	\frac{d\mathbf{S}_i}{dt} = -\gamma\,\mathbf{S}_i\times \bigl(\mathbf{H}_i+\bm{\eta}_i\bigr)
	-\alpha\, \mathbf{S}_i\times \bigl[ \mathbf{S}_i\times \bigl(\mathbf{H}_i+\bm{\eta}_i\bigr) \bigr],
\end{equation}
where $\gamma$ is the gyromagnetic ratio, $\alpha$ is the damping parameter, and $\bm{\eta}_i$ denotes a stochastic magnetic field describing thermal fluctuations. The deterministic effective field $\mathbf{H}_i$ is obtained directly from the learned energy functional through automatic differentiation, while the stochastic contribution is generated according to the fluctuation--dissipation theorem.
In practical simulations, the GNN therefore serves as a surrogate magnetic force field. Given an instantaneous spin configuration, the network predicts the total energy $E$, from which the effective fields $\mathbf{H}_i$ are computed by automatic differentiation and subsequently used to evolve the spins according to the stochastic LLG equation. This procedure enables large-scale finite-temperature spin-dynamics simulations without repeatedly solving the underlying electronic problem.

The physically relevant quantity entering the spin dynamics is the torque
\begin{equation}
	\label{eq:torque_def}
	\bm \tau_i = \mathbf{S}_i\times\mathbf{H}_i,
\end{equation}
which determines both the precessional motion and the dissipative relaxation of the local moments. In contrast, the component of the effective field parallel to $\mathbf{S}_i$ does not contribute to the equation of motion and is therefore dynamically irrelevant. Consequently, accurate prediction of the effective field does not necessarily imply accurate prediction of the resulting spin dynamics. For machine-learning force fields intended for spin-dynamics simulations, the torque thus provides a more physically meaningful measure of model fidelity than the effective field itself.

Motivated by this observation, the network parameters are optimized by minimizing the mean-squared error (MSE) between the reference torques and those predicted by the GNN,
\begin{equation}
	\label{eq:loss_func}
	\mathcal{L} = \frac{1}{N} \sum_i \left| \bm \tau_i^{\,\mathrm{ML}} - \bm \tau_i^{\,\mathrm{exact}} \right|^2,
\end{equation}
where $N$ denotes the number of lattice sites. Since the torques are obtained from effective fields derived through automatic differentiation of the learned energy functional, minimizing the torque error simultaneously constrains the underlying energy landscape that governs the spin dynamics. As demonstrated below, this training strategy yields highly accurate torques and enables faithful reproduction of the nonequilibrium spin dynamics generated by the underlying electronic model.

\section{Validation of the GNN Force Field}
\label{sec:validation}

\begin{figure*}[t]
\centering
\includegraphics[width=1.99\columnwidth]{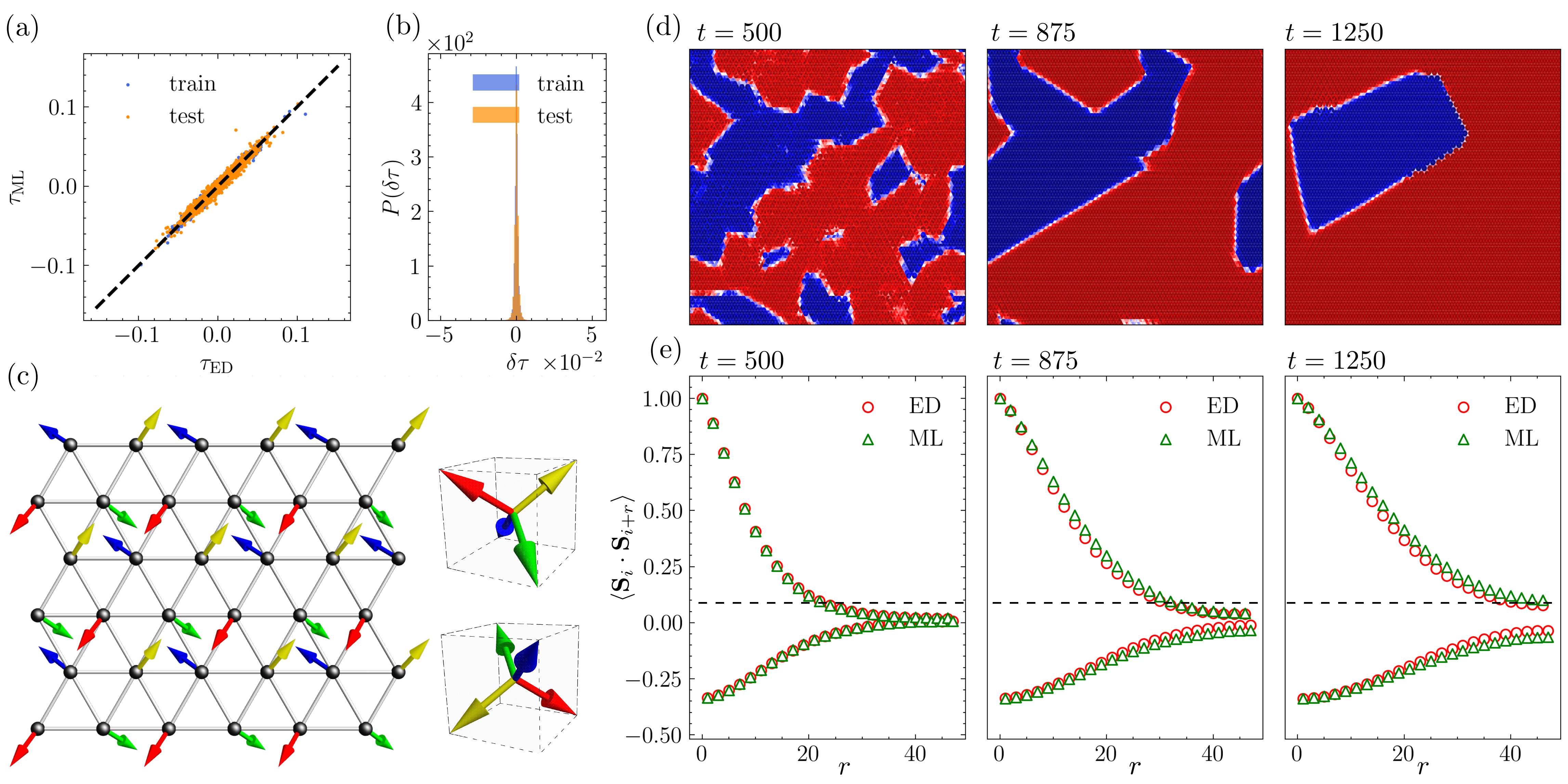}
\caption{
Benchmark of the graph-neural-network force field for the noncoplanar tetrahedral state on the triangular lattice. (a) Comparison of torques predicted by the GNN and exact electronic calculations. (b) Corresponding distribution of torque errors. (c) Schematic illustration of the tetrahedral spin order and the two degenerate states with opposite chirality. (d) Representative snapshots of the local scalar chirality $\chi_{ijk}=\mathbf S_i\cdot(\mathbf S_j\times\mathbf S_k)$, revealing the growth and coarsening of chiral domains associated with the discrete $Z_2$ order parameter. (e) Equal-time spin-spin correlation functions at selected times, comparing direct KPM--LLG simulations with GNN--LLG results. The excellent agreement demonstrates that the learned force field accurately reproduces both the local torques and the nonequilibrium dynamics of the chiral tetrahedral phase.
}    \label{fig:tetra-benchmark}
\end{figure*}

\subsection{Itinerant \texorpdfstring{$s$--$d$}{s-d} model and training data}

To demonstrate the performance of the proposed graph-neural-network force field, we consider the standard one-band $s$--$d$ model~\cite{Zener1951a,deGennes1960,anderson61}, which describes itinerant electrons coupled to localized classical moments. The Hamiltonian is
\begin{equation}
	\mathcal{H} = -\sum_{\langle ij\rangle,\alpha} t_{ij} \left( c_{i\alpha}^{\dagger} c_{j\alpha} + \mathrm{H.c.} \right)
	- J \sum_i \sum_{\alpha\beta} c_{i\alpha}^{\dagger} \, \bm{\sigma}_{\alpha\beta} \, c_{i\beta} \cdot \mathbf S_i ,
	\label{eq:sd}
\end{equation}
where $c_{i\alpha}^{\dagger}$ creates an itinerant electron with spin index $\alpha$ on lattice site $i$, $t_{ij}$ denotes the hopping amplitude, $J$ is the local Hund's-rule coupling, and $\bm{\sigma}=(\sigma^x,\sigma^y,\sigma^z)$ is the vector of Pauli matrices. The first term describes the kinetic motion of conduction electrons, while the second term couples the electronic spin density to the localized magnetic moments. Despite its apparent simplicity, the model exhibits a rich variety of electronically generated magnetic phases arising from the competition between electron delocalization and local spin alignment.

The spin dynamics is governed by the effective magnetic fields generated by the itinerant electrons. Applying the Hellmann--Feynman theorem to the electronic free energy yields the local field $\mathbf H_i = J \sum_{\alpha\beta}\bm{\sigma}_{\alpha\beta}\, C_{i\beta,i\alpha}$, where $C_{i\beta,j\alpha} = \langle c_{j\alpha}^{\dagger} c_{i\beta} \rangle$ denotes the single-particle density matrix. The central computational challenge is therefore the evaluation of the electronic correlation matrix $C$ for an arbitrary spin configuration $\{\mathbf S_i\}$. In the present work, the reference data are generated from direct electronic calculations of the underlying Hamiltonian, using either exact diagonalization (ED) or the kernel polynomial method (KPM)~\cite{weisse2006,barros2013,wang2018}, the latter enabling the larger system sizes required to capture the domain structures and coarsening dynamics of complex chiral magnetic phases. Because the electronic spectrum and density matrix depend sensitively on the instantaneous spin background, they must be recomputed repeatedly throughout the Landau--Lifshitz evolution. Consequently, direct spin-dynamics simulations require the repeated solution of a large quantum-mechanical problem, making long-time simulations of large systems computationally prohibitive.

The objective of the GNN force field is to learn the mapping between local magnetic environments and the electronically generated torques without explicitly solving the electronic problem. To this end, the training data are generated from exact electronic calculations of the $s$--$d$ model. For each spin configuration, the electronic Hamiltonian is solved to obtain the electronic density matrix, from which the effective fields and torques are computed using the Hellmann--Feynman theorem. The resulting dataset therefore provides a direct correspondence between local magnetic environments and the electronically generated torques that drive the spin dynamics. 

Once trained, the GNN replaces the repeated electronic calculation with a rapid surrogate model that predicts the magnetic energy functional and the corresponding torques directly from the spin configuration. Unlike direct electronic calculations, whose computational cost is dominated by repeatedly solving the electronic Hamiltonian during the time evolution, the trained GNN evaluates the magnetic energy functional through a sequence of inexpensive message-passing operations. This reduction in computational complexity enables simulations on substantially larger spatial and temporal scales.

\subsection{Benchmarking across distinct magnetic textures}

To assess both the accuracy and the transferability of the proposed force field, we benchmark the model at three increasingly demanding levels: local torque prediction, nonequilibrium spin dynamics, and emergent collective ordering kinetics. We consider three representative magnetic states of increasing complexity: a collinear N\'eel antiferromagnet on the square lattice, a coplanar $120^\circ$ state on the triangular lattice, and a noncoplanar tetrahedral state on the triangular lattice. These states span three qualitatively distinct classes of magnetic textures commonly encountered in metallic magnets and therefore provide increasingly stringent tests of the ability of the learned force field to capture electronically mediated spin interactions. From the perspective of spin-dynamics simulations, they probe the performance of the GNN across a broad range of magnetic environments, ranging from simple bipartite antiferromagnets to frustrated magnetic states with finite scalar chirality.

The benchmark datasets were generated from direct electronic calculations of the $s$--$d$ model and encompass a wide range of spin configurations sampled during nonequilibrium relaxation dynamics. The N\'eel benchmark consists of approximately $8.1\times10^4$ spin configurations generated on a $24\times24$ square lattice, while the $120^\circ$ benchmark contains $5.0\times10^4$ configurations on a $24\times24$ triangular lattice. The tetrahedral benchmark, which represents the most demanding test considered here, contains more than $10^4$ configurations generated on a substantially larger $96\times96$ lattice. Together, these datasets provide a comprehensive test of whether the graph-based magnetic force-field framework can accurately reproduce electronically generated torques and long-time spin dynamics across diverse magnetic textures.

We first examine the accuracy of the torque prediction. The parity plots shown in Figs.~2(a), 3(a), and 4(a) demonstrate excellent agreement between the torques predicted by the GNN and those obtained from direct electronic calculations. For the square-lattice N\'eel state, the trained model achieves a test-set coefficient of determination $R^2 = 0.9995$, a mean-squared error (MSE) of $5.25\times10^{-6}$, and a mean absolute error (MAE) of $1.50\times10^{-3}$. For the triangular-lattice $120^\circ$ state, the corresponding metrics are $R^2 = 0.9982$, MSE $= 8.42\times10^{-6}$, and MAE $= 1.82\times10^{-3}$. Even for the noncoplanar tetrahedral state, which exhibits the most complex magnetic environment considered in this work, the model achieves $R^2 = 0.9589$, MSE $= 1.79\times10^{-6}$, and MAE $= 8.87\times10^{-4}$. The corresponding error distributions, shown in Figs.~2(b), 3(b), and 4(b), remain sharply centered around zero with minimal systematic bias. Remarkably, comparable predictive accuracy is achieved across collinear, coplanar, and noncoplanar magnetic environments despite their substantially different local spin correlations. These results demonstrate that the combination of symmetry-preserving descriptors and hierarchical graph message passing provides a sufficiently expressive representation to accurately learn electronically generated torque landscapes across a broad range of magnetic environments.

\begin{figure}[t]
\centering
\includegraphics[width=0.99\columnwidth]{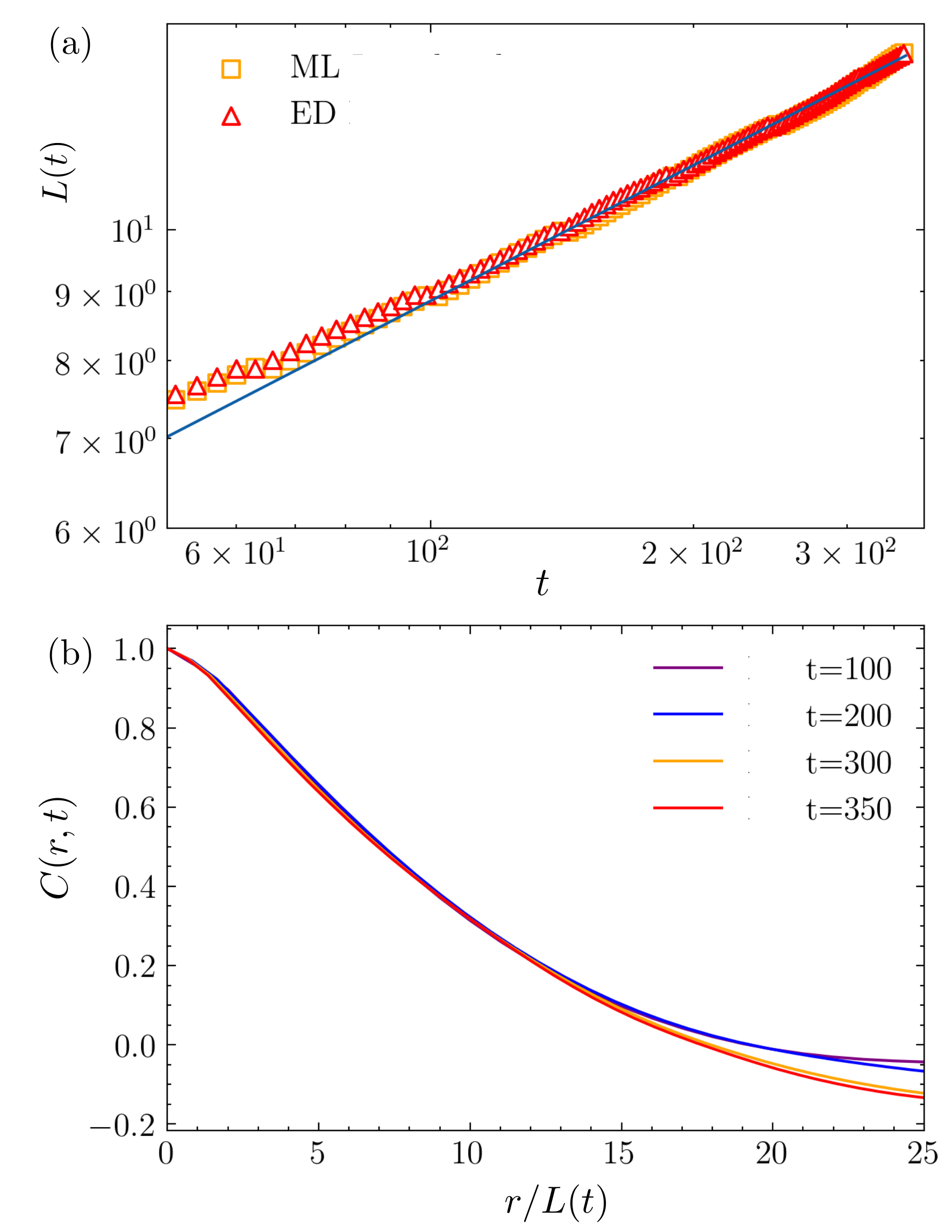}
\caption{Further benchmark of the GNN force field for the noncoplanar tetrahedral phase. (a) Characteristic chiral-domain size $L(t)$ as a function of time obtained from direct electronic simulations and GNN-based spin dynamics. The learned force field accurately reproduces the previously reported approximately linear coarsening law $L(t)\sim t$. (b) Chirality correlation function $C(r,t)$ plotted against the scaled distance $r/L(t)$. The collapse of data at different times demonstrates dynamical scaling during chiral-domain coarsening. Together, these results show that the GNN force field faithfully captures the large-scale nonequilibrium dynamics of the chiral phase.
}    \label{fig:L-vs-t}
\end{figure}

While accurate torque prediction is a necessary requirement for a magnetic force field, the ultimate objective is the faithful reproduction of nonequilibrium spin dynamics. We therefore benchmark the GNN-generated dynamics against direct electronic simulations. Figures~2(d) and 3(d) show representative spin configurations obtained during relaxation toward the N\'eel and $120^\circ$ ordered states, respectively, following a thermal quench from a disordered initial condition. In both cases, the spin configurations progressively develop the characteristic antiferromagnetic correlations associated with the corresponding ordering wave vectors. For the tetrahedral phase, the relevant order parameter is the local scalar chirality, and representative snapshots of the chirality field are shown in Fig.~4(d). Since the tetrahedral state possesses a discrete $Z_2$ chirality degree of freedom, the relaxation dynamics proceeds through the formation and subsequent coarsening of chiral domains. In all three cases, the GNN accurately reproduces the characteristic evolution of the magnetic texture throughout the ordering process.

A complementary benchmark is provided by collective observables. For the square-lattice antiferromagnet, Fig.~2(c) compares the time dependence of the spin structure factor $\mathcal{S}(\pi,\pi)$, while Fig.~3(c) shows the corresponding structure factor evaluated at the $K$ point of the hexagonal Brillouin zone for the triangular-lattice $120^\circ$ state. In both cases, the GNN-generated dynamics closely tracks the direct electronic results throughout the entire relaxation process. The agreement demonstrates that the learned force field accurately reproduces not only instantaneous local torques, but also the collective ordering kinetics that emerges from their long-time evolution.

Figures~2(e), 3(e), and 4(e) further compare the equal-time spin-spin correlation functions $\langle \mathbf S_i \cdot \mathbf S_{i+r} \rangle$ obtained from GNN and direct electronic simulations at representative times during the coarsening process. The correlation functions are averaged over many independent initial conditions to reduce statistical fluctuations. Across all three benchmark systems, the learned force field faithfully reproduces both short-range magnetic correlations and the growth of long-range correlation lengths. The agreement remains robust despite the substantial differences in magnetic symmetry, ordering patterns, and underlying electronic mechanisms among the three phases. Together with the structure-factor benchmarks, these results indicate that the learned energy landscape remains sufficiently accurate under repeated differentiation and time integration to support stable long-time spin-dynamics simulations.

\subsection{Chiral-domain coarsening}

Beyond conventional correlation-based benchmarks, the tetrahedral phase provides an even more stringent test of the learned dynamics. As shown in Fig.~4(d), the coarsening process is characterized by an unusual domain morphology consisting of extended, nearly straight interfaces connected by sharp corners and vertices. Such faceted chiral domains originate from the strong orientational dependence of the domain-wall energy induced by the underlying electronic structure~\cite{Ozawa2017}. Figure~\ref{fig:L-vs-t}(a) compares the characteristic chiral-domain size $L(t)$ obtained from direct electronic simulations and GNN-based dynamics. The learned force field accurately reproduces the previously reported approximately linear growth law $L(t)\sim t$~\cite{Fan24}. This behavior lies outside the conventional Allen--Cahn picture, in which the local velocity of a domain wall is proportional to its curvature. 

Since the faceted interfaces observed in Fig.~4(d) are nearly straight, their local curvature is close to zero, implying that conventional curvature-driven growth would be strongly suppressed. Instead, the coarsening dynamics is governed by the evolution and annihilation of the sharp vertices where multiple faceted interfaces meet. These vertices behave as relatively stable point-like structural defects whose dynamics controls the growth of the characteristic domain size. Consistent with this picture, Fig.~\ref{fig:L-vs-t}(b) shows that chirality correlation functions measured at different times collapse onto a universal curve when plotted as a function of the scaled distance $r/L(t)$, demonstrating dynamical scaling during the coarsening process. The excellent agreement between the GNN and direct electronic simulations demonstrates that the learned force field faithfully captures not only the local torque landscape and magnetic correlations, but also the emergent large-scale kinetics governing chiral-domain evolution.

These benchmarks establish the proposed GNN as a transferable magnetic force-field framework capable of reproducing both local torque landscapes and emergent nonequilibrium spin dynamics across a diverse set of itinerant magnetic states. More importantly, they demonstrate that hierarchical graph representations constructed from local symmetry-preserving descriptors provide an effective route for learning the long-ranged, nonlinear interactions generated by itinerant electrons.

\section{Summary and Outlook}

In this work, we have demonstrated that graph neural networks can serve as accurate magnetic force fields for itinerant magnets, enabling large-scale nonequilibrium spin-dynamics simulations without repeated solution of the underlying electronic problem. Analogous to the role played by machine-learned interatomic potentials in atomistic simulations, the present framework replaces expensive electronic calculations with a learned energy functional from which effective magnetic fields and torques are obtained through automatic differentiation. The resulting force field accurately reproduces both instantaneous torques and emergent dynamical evolution across a diverse set of magnetic textures, including collinear, noncollinear, and noncoplanar states. These results establish graph-based magnetic force fields as a promising computational framework for studying electronically driven spin dynamics over length and time scales that are inaccessible to direct electronic simulations.

The present work focuses on itinerant magnetic systems whose dominant interactions originate from electronically mediated exchange processes and which approximately preserve global spin-rotation symmetry. While this approximation neglects relativistic spin-orbit effects, it nevertheless describes a broad class of metallic magnetic materials. Indeed, variants of the $s$--$d$ model provide the standard theoretical framework for understanding metallic ferromagnets, frustrated itinerant magnets, chiral magnetic textures, and a wide range of spintronic phenomena. Many of the essential collective processes of interest in these systems, including domain formation, coarsening, and the dynamics of complex magnetic textures, are governed primarily by exchange interactions and can therefore be captured within the present symmetry-preserving framework. From this perspective, the benchmark systems considered here should not be viewed merely as model Hamiltonians, but rather as representative realizations of electronically driven magnetism in metallic materials.

At the same time, an important limitation of the current approach is the assumption that spin and spatial degrees of freedom can be treated separately, corresponding to a decoupled $\mathrm{SO}(3)$ spin symmetry and lattice point-group symmetry. In many technologically relevant magnetic materials, spin-orbit coupling generates magnetic anisotropies, Dzyaloshinskii--Moriya interactions, magnetocrystalline effects, and other symmetry-breaking terms that play a central role in determining magnetic structure and dynamics. Extending machine-learned magnetic force fields to such systems therefore represents an important next step toward realistic materials applications.

From a machine-learning perspective, incorporating spin-orbit coupling poses a qualitatively different challenge from the present work. In the absence of spin-orbit interactions, spin and spatial degrees of freedom transform independently under global spin rotations and lattice symmetry operations. Once spin-orbit coupling is introduced, these transformations become intrinsically intertwined, requiring neural-network architectures that simultaneously respect both spin and spatial symmetries. We anticipate that such systems will require a nontrivial integration of equivariant neural-network techniques with graph-based message-passing architectures, combining the symmetry-aware representations of equivariant models with the flexibility and scalability of modern graph neural networks. Developing such equivariant magnetic force fields constitutes a particularly promising direction for future research.

Beyond spin-orbit coupling, another important avenue is the extension to multi-orbital itinerant-electron models, where orbital degrees of freedom, crystal-field effects, and orbital-selective correlations can generate substantially richer magnetic interactions. Because the present framework is formulated directly in terms of local magnetic environments and graph-based message passing, the underlying methodology is sufficiently general that it can be adapted to more realistic microscopic descriptions while retaining the same force-field philosophy.

More broadly, the results presented here suggest that machine-learned magnetic force fields may provide a new route toward multiscale simulations of complex magnetic materials. By bridging microscopic electronic models and large-scale spin-dynamics simulations, such approaches offer the possibility of studying collective magnetic phenomena across previously inaccessible spatial and temporal scales. We hope that the present work serves as a step toward a broader computational framework in which machine learning enables predictive dynamical simulations of electronically driven magnetism in realistic materials.

\begin{acknowledgments}
This work was supported by the US Department of Energy Basic Energy Sciences under Contract No. DE-SC0020330. The authors acknowledge Research Computing at The University of Virginia for providing computational resources and technical support that have contributed to the results reported within this publication. 
\end{acknowledgments}

\appendix

\section{Reference electronic calculations}

\begin{table*}
\centering
\caption{GNN architecture and training hyperparameters used for the three magnetic benchmark systems.}
\label{tab:gnn_hyper}
\begin{tabular}{lccc}
\hline\hline
 & N\'eel & $120^\circ$ & Tetrahedral \\
\hline

Node feature sizes &
[1024, 738, 512, 512, 256] &
[1024, 738, 738, 512, 512, 256] &
[1280, 1024, 738, 512, 512, 256] \\

Edge feature sizes &
[1, 256, 256, 128, 128] &
[1, 256, 256, 256, 128, 128] &
[1, 512, 256, 184, 128, 128] \\

Readout MLP &
[256, 128, 64, 16, 1] &
[256, 128, 64, 16, 1] &
[256, 128, 64, 16, 1] \\

Activation &
GELU &
GELU &
GELU \\

Aggregation scheme &
Weighted message passing &
Weighted message passing &
Weighted message passing \\

Optimizer &
AdamW &
AdamW &
AdamW \\

Learning rate &
$5\times10^{-4}$ &
$5\times10^{-4}$ &
$3\times10^{-4}$ \\

Weight decay &
$10^{-3}$ &
$10^{-3}$ &
$10^{-3}$ \\

Scheduler &
Cosine annealing &
Cosine annealing &
Cosine annealing \\

Batch size &
8 &
8 &
6 \\

Training epochs &
100 &
100 &
120 \\

Trainable parameters &
4,389,579 &
5,993,398 &
8,444,836 \\

Train/test split &
80/20 &
80/20 &
80/20 \\

\hline\hline
\end{tabular}
\end{table*}

The training labels for the GNN force field were generated from the $s$--$d$ Hamiltonian introduced in Eq.~(\ref{eq:sd}). Within the adiabatic approximation, the electronic degrees of freedom are assumed to remain in equilibrium for a given spin configuration $\{\mathbf S_i\}$, thereby generating an effective free-energy landscape for the localized moments. For a fixed spin background, the $s$--$d$ Hamiltonian is bilinear in the fermionic operators and can therefore be represented by a single-particle Hamiltonian matrix $\mathbb H$ with matrix elements
\begin{equation}
	\mathbb H_{i\alpha,j\beta} = -t_{ij}\delta_{\alpha\beta} - J\,\delta_{ij}\, \mathbf S_i\cdot \boldsymbol{\sigma}_{\alpha\beta},
	\label{eq:first_quantized_H}
\end{equation}
where $i,j$ denote lattice sites and $\alpha,\beta$ are spin indices. The corresponding second-quantized Hamiltonian can be written as $\mathcal H = \sum_{i\alpha,j\beta} \mathbb H_{i\alpha,j\beta}\, c^\dagger_{i\alpha} c^{\,}_{j\beta}$, so that the electronic problem is completely specified by the single-particle Hamiltonian matrix $\mathbb H$. Once the electronic density matrix is obtained, the effective magnetic field acting on the localized moments follows from the Hellmann--Feynman theorem,
\begin{equation}
	\mathbf H_i = - \Bigl\langle \frac{\partial \mathcal H}{\partial \mathbf S_i} \Bigr\rangle = J \sum_{\alpha\beta} \boldsymbol{\sigma}_{\alpha\beta} C_{i\beta,i\alpha},
\end{equation}
where
\begin{equation}
	C_{i\beta,j\alpha} = \bigl\langle c^\dagger_{j\alpha} c^{\,}_{i\beta} \bigr\rangle
\end{equation}
is the single-particle density matrix. The corresponding torque is then obtained from Eq.~(\ref{eq:torque_def}).

For the square-lattice N\'eel and triangular-lattice $120^\circ$ benchmarks, the density matrix was computed using exact diagonalization (ED) of the Hamiltonian matrix $\mathbb H$. Denoting the eigenvalues and eigenvectors by $\epsilon_n$ and $U^{(n)}_{i,\alpha}$, respectively, the density matrix is given by
\begin{equation}
	C_{i\beta,j\alpha} = \sum_n f(\epsilon_n)\, U^{(n)}_{i,\beta} U^{(n)*}_{j,\alpha},
\end{equation}
where $f(\epsilon)$ is the Fermi--Dirac distribution function. The resulting effective fields and torques provide the reference labels used for training the GNN.

For the tetrahedral benchmark, which involves lattices as large as $96\times96$, explicit diagonalization becomes computationally prohibitive because the computational cost scales cubically with the matrix dimension. In this case, the density matrix was evaluated using the kernel polynomial method (KPM)~\cite{weisse2006}, which avoids explicit computation of the electronic eigenspectrum. The central idea of KPM is to approximate the electronic free-energy functional and related matrix functions through a finite expansion in Chebyshev polynomials of a suitably rescaled Hamiltonian matrix.

Starting from a random vector, the action of successive Chebyshev polynomials of the Hamiltonian generates a sequence of Chebyshev vectors $\{\boldsymbol{\alpha}^{(m)}\}$ through recursive sparse matrix--vector multiplications. A second sequence of vectors $\{\boldsymbol{\beta}^{(m)}\}$ is subsequently constructed through a reverse accumulation procedure. Together, these vectors provide an efficient representation of the electronic density matrix without requiring explicit computation of the electronic eigenstates. The density matrix can be reconstructed as~\cite{barros2013,wang2018}
\begin{equation}
	C_{i\beta,j\alpha} = \beta^{(0)}_{i\beta}\alpha^{(0)}_{j\alpha} + 2\sum_{m=1}^{M-2} \beta^{(m)}_{i\beta} \alpha^{(m)}_{j\alpha},
\end{equation}
where $M$ is the truncation order of the Chebyshev expansion. Importantly, both the forward and backward recursions involve only sparse matrix--vector multiplications and therefore scale linearly with system size. This favorable scaling enables efficient computation of electronically mediated torques on lattices substantially larger than those accessible to exact diagonalization while maintaining controlled numerical accuracy. Additional details of the KPM implementation can be found in Ref.~\cite{barros2013,wang2018}.

\section{GNN Implementation Details}

\begin{table*}
\centering
\caption{Summary of the training datasets used for the three magnetic benchmark systems.}
\label{tab:datasets}
\begin{tabular}{lccc}
\hline\hline
 & N\'eel & $120^\circ$ & Tetrahedral \\
\hline

Lattice &
$24\times24$ &
$24\times24$ &
$96\times96$ \\

Electronic solver &
ED &
ED &
KPM \\

Number of graphs &
81,000 &
50,337 &
10,530 \\

Damping $\alpha$ &
0.1 &
0.1 &
0.075 \\

Time step $\Delta t$ &
0.005 &
0.02 &
0.025 \\

Trajectory length &
35,000 steps &
50,000 steps &
50,000 steps \\

Sampling method &
Relaxation trajectories &
Relaxation trajectories &
30 relaxation ensembles \\

Additional sampling &
Random configurations &
Random configurations &
Snapshots every 100 steps \\

\hline\hline
\end{tabular}
\end{table*}

The graph-neural-network architecture follows the message-passing framework introduced in Sec.~II. Lattice sites are represented as graph nodes, while bonds connecting neighboring sites are represented as graph edges. The initial node and edge features are constructed from the symmetry-preserving spin descriptors described in Sec.~II, including bond correlations, local scalar chiralities, and triangle-based invariants. These descriptors provide an invariant representation of the local magnetic environment while preserving the global spin-rotation symmetry and lattice symmetries of the underlying spin model.

The node and edge features are subsequently refined through a sequence of message-passing layers. At each layer, information from neighboring nodes and edges is aggregated over distinct coordination shells using trainable weight matrices associated with each shell. This shell-dependent aggregation allows the network to learn the distance dependence of electron-mediated spin interactions while maintaining the symmetry equivalence of sites within the same coordination shell. Simultaneous updates of node and edge features enable the network to capture both local magnetic environments and higher-order many-body spin correlations generated by itinerant electrons.

Following the final message-passing layer, the node features are processed by a multilayer perceptron (MLP) that outputs a local energy contribution for each lattice site. The total energy is obtained by summing these site-resolved contributions over all nodes. This extensive energy construction ensures the correct scaling with system size and allows the trained model to be transferred to lattices substantially larger than those used during training.

The effective magnetic field acting on each spin is then obtained through automatic differentiation of the predicted total energy with respect to the spin degrees of freedom. The resulting torque enters directly into the Landau--Lifshitz--Gilbert dynamics. Because the force field is derived from a single scalar energy functional, the learned dynamics remains internally consistent and respects the continuous spin-rotation symmetry of the underlying Hamiltonian.

All models were implemented using PyTorch and PyTorch Geometric. The architecture and training hyperparameters used for the three benchmark systems are summarized in Table~\ref{tab:gnn_hyper}. Although the overall message-passing framework remains identical across all benchmarks, the node-feature dimensions, edge-feature dimensions, and training parameters were optimized separately for each magnetic phase.

\section{Training datasets and optimization}

The network was trained using the torque-based loss function defined in Eq.~(\ref{eq:loss_func}). Although the GNN predicts the total energy of a spin configuration, the physically relevant quantity governing the spin dynamics is the local torque acting on each spin. Since the effective magnetic field is obtained through differentiation of the learned energy functional, small energy errors can potentially lead to amplified force errors. Direct minimization of the torque error therefore provides a more stringent and physically meaningful training objective, ensuring accurate reproduction of the electronic forces responsible for the spin dynamics.

Training configurations were generated from stochastic spin-dynamics simulations of the underlying $s$--$d$ model. To ensure broad coverage of configuration space, the datasets were constructed from a combination of relaxation trajectories and additional nonequilibrium spin configurations. Consequently, the training data include highly disordered states, partially ordered configurations encountered during relaxation, and configurations close to the equilibrium ordered phases. This diverse sampling strategy improves the robustness and transferability of the learned force field during long-time dynamical simulations.

For the N\'eel and $120^\circ$ benchmarks, the reference torques were generated using exact diagonalization (ED) of the electronic Hamiltonian on $24\times24$ lattices. For the tetrahedral benchmark, where substantially larger systems were required to capture the coarsening dynamics of chiral domains, the electronic forces were computed using the kernel polynomial method (KPM) on a $96\times96$ lattice. Depending on the benchmark, the resulting datasets contain between approximately $10^4$ and $10^5$ graph configurations spanning collinear, noncollinear, and noncoplanar spin textures. Details of the dataset generation procedures are summarized in Table~\ref{tab:datasets}, while the corresponding GNN architectures and optimization parameters are listed in Table~\ref{tab:gnn_hyper}.

Optimization was performed using the AdamW optimizer together with cosine-annealing learning-rate scheduling. An 80/20 train--test split was employed for all datasets. The final models were selected according to the lowest validation loss and subsequently used in the large-scale GNN--LLG simulations discussed in Sec.~III.

\bibliography{ref.bib}

\end{document}